\documentclass[useAMS,usenatbib]{mn2e}

\usepackage{graphicx}
\usepackage{xfrac}
\title[New RR Lyrae variables in binary systems]{New RR Lyrae variables in binary systems}
\author[G. Hajdu et al.]
{G. Hajdu$^{1,2}$\thanks{E-mail:
ghajdu@astro.puc.cl; mcatelan@astro.puc.cl},
M. Catelan$^{1,2}$,
J. Jurcsik$^{3}$,
I. D\'ek\'any$^{2,1}$,
A. J. Drake$^{4}$ and
J.-B. Marquette$^{5}$\\
$^{1}$Instituto de Astrof\'{i}sica, Facultad de F\'{i}sica, Pontificia Universidad
Cat\'olica de Chile, Av. Vicu\~na Mackenna 4860, 782-0436 Macul, Santiago, Chile\\
$^{2}$Millennium Institute of Astrophysics, Santiago, Chile\\
$^{3}$Konkoly Observatory, H-1525 Budapest, PO Box 67, Hungary\\
$^{4}$California Institute of Technology, 1200 East California Boulevard, CA 91225, USA\\
$^{5}$Institut d'Astrophysique de Paris, Universit\'e Pierre et Marie Curie, CNRS UMR7095, 98 bis Boulevard Arago, F-75014 Paris, France
}
\begin{document}

\date{Accepted xxx. Received yyy; in original form zzz}

\pagerange{\pageref{firstpage}--\pageref{lastpage}} \pubyear{2015}

\maketitle

\label{firstpage}

\begin{abstract}
Despite their importance, very few RR~Lyrae (RRL) stars have been 
known to reside in binary systems. 
We report on a search for binary RRL in the OGLE-III Galactic bulge data. 
Our approach consists in the search for evidence of the light-travel 
time effect in so-called {\em observed minus calculated} ($O-C$) diagrams.
Analysis of 1952 well-observed fundamental-mode RRL in the OGLE-III data revealed 
an initial sample of 29 candidates. We used the recently released OGLE-IV 
data to extend the baselines up to 17 years, leading to a final sample of 
12 firm binary candidates. We provide $O-C$ diagrams and binary parameters for 
this final sample, and also discuss the properties of 8 additional candidate 
binaries whose parameters cannot be firmly determined at present. 
We also estimate that $\ga4$ per cent of the RRL reside in binary systems.
\end{abstract}


\section{Introduction}

\label{intro}

RR~Lyrae (RRL) stars play a key role in astrophysics. They are important distance indicators,
allowing us to determine the distance to the
closest galaxies \citep[e.g.,][]{cc13,da13}, and thus providing an important step in
the calibration of the extragalactic distance scale. Their importance in the context of 
galaxy formation and evolution is also being increasingly recognized \citep[e.g.,][]{mc09}. Indeed, 
RRL stars being unmistakably old, 
their distribution in the Galactic halo provides evidence of the early Milky Way formation history
\citep{ajdea13a,ajdea13b,se13,gtea14}. In addition, 
RRL stars help trace the spatial distribution, 
and even the age, of some of the Milky Way's oldest stellar populations \citep{ywl92,cdf93,de13}.

Despite their usefulness, there are still important uncertainties affecting the fundamental parameters  
and physical properties of these stars. Trigonometric parallaxes of even the closest 
RRL are notorious for their large error bars \citep{fbea11}. To complicate matters further, 
the exact evolutionary status of even RR~Lyr itself is uncertain, with indications that it 
may be significantly overluminous, compared to the zero-age horizontal branch (ZAHB) level 
at its metallicity \citep{cc08,mfea08}. 

RRL variables occupy a fairly narrow 
strip~-- the so-called {\em instability strip}~-- at intermediate temperatures along the HB. As such, 
their exact mass value is crucial in establishing whether an HB star will ever become an RRL, 
or instead, a non-variable blue or red HB star. In this sense, theory predicts that the
masses of RRL stars should decrease with increasing metallicity, with little scatter at any 
given [Fe/H] \citep[e.g.,][]{mc92,as06}. Direct empirical confirmation 
of this important result is, however, still lacking. Most of the available information regarding 
RRL masses comes from the so-called {\em Petersen diagram} \citep{jop73} of double-mode 
RRL (RRd) stars. RRd stars are observed to pulsate simultaneously in the fundamental 
and first overtone radial modes, and their distribution in the period ratio vs. period (Petersen)
diagram is predicted to be a strong function of the pulsating star's mass, in addition to other 
parameters, such as metallicity \citep[e.g.,][]{bpea00}. 
RRL star masses can thus be derived by comparing the observed positions of RRd stars in 
the Petersen diagram with those predicted according to stellar evolution and pulsation theory 
\citep[e.g.,][]{gbea96,de08}. However, this method can only be trusted to provide accurate 
masses if the theoretical framework upon which it is based is itself accurate. To constrain the 
theories themselves, it is imperative to obtain a model-independent mass measurement.

Binary systems allow the derivation of the masses of its components, if the orbital
parameters are known. In the case of binary systems containing classical 
Cepheids, analysis of their orbital parameters has played a crucial role in accurately 
establishing their physical properties, including their masses \citep{gpea10}. 
Furthermore, Cepheid-bearing binary systems are relatively common \citep{sza03}, 
and even systems in which {\em both} components are Cepheids are now known 
to exist \citep{wgea14}.

The situation regarding RRL stars could hardly be more different, as for long only one 
RRL, TU~UMa, has been known to reside in a binary system 
(see \citealt{sw90} and \citealt{wa99}, for very detailed analyses of this star).
Recently, eclipsing binary RRL candidates have been found by the OGLE project
in the Galactic bulge \citep{so11}, as well as in the LMC \citep{so09}. However, follow-up 
observations and modeling of the bulge candidate have shown that the pulsating component 
of this system has too low a mass to be a bona-fide RRL star \citep{pi12,sm13}. 
Other candidate eclipsing 
RRL found by the OGLE-II project in the LMC \citep{so03} have turned out to be optical 
blends \citep{pr08}. Careful analyses of the RRL LCs of the \textit{Kepler}
mission have uncovered only three possible binary candidates \citep{lq14,gs14}.

Recently, a number of metal-poor, carbon-rich RRLs have been identified via spectroscopy
(e.g., \citealt{ke14}). The anomalous C abundance of these stars can 
be explained by mass accretion from a more massive companion, which has evolved through
the asymptotic giant phase \citep{st13}, suggesting that these RRLs might be members of binary
systems in which the other component has already evolved to the white dwarf stage. However, 
the binary nature of these RRLs has not yet been directly established.

In this {\em Letter}, we describe our search for periodic phase variations caused by the light-travel time 
effect \citep{ir52} in the LCs of a subsample of well-observed RRL stars from the 
OGLE-III survey towards the Galactic bulge \citep{so11}. We augment the LCs of the 
binary candidates thus obtained with newly published photometry from the OGLE-IV survey 
\citep{so14}, in order to increase the observational baseline.
We derive binary parameters for the best candidates. Furthermore, we discuss the RRL binary fraction,
as well as the detectability of such systems through the light-travel time effect.

\begin{figure}
\includegraphics[angle=0,scale=.95]{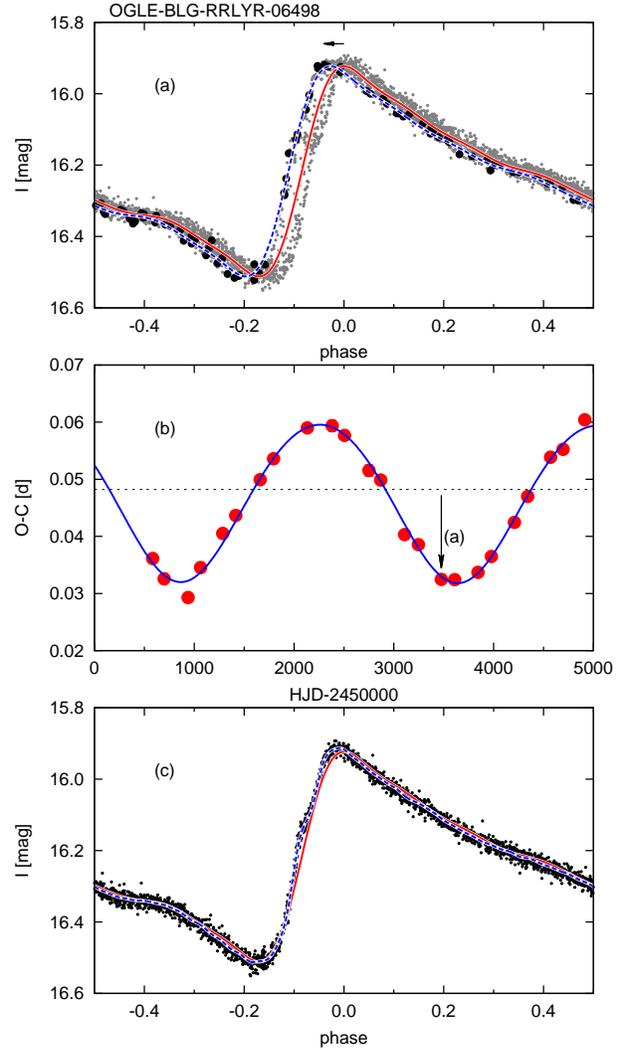}
\caption{$O-C$ analysis of the OGLE-III observations.
(a) The original LC folded with the pulsation period ({\em grey dots}), its Fourier fit ({\em red line}),
one section of the LC ({\em black dots}), and the least-squares template fit ({\em blue dashed line}).
(b) The resulting $O-C$ diagram ({\em dots}), and the corresponding binary-model fit ({\em blue line}). 
The point corresponding to the measurement shown in (b) is indicated by an arrow. 
(c) The folded LC, after correcting the times of observations for the binary motion (panel [b]).
The Fourier fit ({\em blue dashed line}) now follows the real LC shape
much more closely than the original fit ({\em red line}).
\label{method}}
\end{figure}

\section[]{Data and analysis}
\label{dataanalysis}

Our initial analysis is based on the $I$-band OGLE-III LCs for bulge RRL  
\citep{so11}. We analyze stars with
fundamental-mode pulsation (RRab subtype), as well as having
an observational baseline covering more than 10~yr.
These criteria define a subsample of
1952 RRab variables. For these, 
we have utilized the so-called {\em observed minus calculated} ($O-C$) diagrams 
(e.g., \citealt{st05}), adopting a linear ephemeris ($C$) for the variables:

\begin{equation}
C(t)=t_0+P_{\mathrm{puls}} \, E,
\end{equation}

\noindent where $t_0$ is the initial epoch, $P_{\mathrm{puls}}$ is the pulsation period,
and $E$ is the epoch number, corresponding to the number of elapsed pulsation cycles since $t_0$.
Times have been transformed into Barycentric Julian Dates (BJD) in the Barycentric Dynamical Time standard 
(\citealt{esg10}).

In the case of variable stars, the $O-C$ diagram is most commonly constructed
by subtracting the ephemeris ($C$) from timing observations ($O$) of particular features of the LCs,
such as maxima or minima, and plotting this quantity as a function of time. 
\citet{he19} proposed utilizing the whole LC for deriving $O-C$
points of a variable by fitting the phase of a LC template to
the observations.
Due to the generally sparse sampling of
OGLE observations, determining the times of individual maxima at different epochs is unfeasible,
making \citeauthor{he19}'s method vastly superior for obtaining the required $O-C$ measurements. 
We thus adopt the latter method in our analysis.

Figure~\ref{method} illustrates the procedure for one of the binary candidates. 
First we created a LC template 
by fitting the original LC with a Fourier series using \textsc{lcfit} \citep{so12},
as shown in Figure~\ref{method}a.
The LCs were divided into short sections corresponding to different observing seasons.
Sections longer than 160 days were split in two, in order to achieve better time resolution.
We have derived the $O-C$ points by least-squares fitting the LC templates in phase to each of these segments. These phase shifts were then used to construct the $O-C$ diagrams (Fig.~\ref{method}b).

\subsection{Selection of binary candidates}
\label{initial}

The light-travel time effect manifests itself as a strictly periodic phase modulation of the LC
of a variable star, caused by its orbital motion around the common center of mass in a
binary system. The change in the observed times of particular features of the LC, 
and consequently in the corresponding $O-C$ values, have the form:

\begin{equation}
z(t)=a \sin i \, \frac{1-e^2}{1+e\cos(\nu)} \, \sin(\nu+\omega), \label{lteffect}
\end{equation}

\noindent where $a$ is the semi-major axis, $i$ is the inclination, $e$ is the eccentricity, and $\omega$ is the
argument of the periastron. The true anomaly $\nu$ is a function of the time $t$,
the orbital period $P_{\rm orb}$, the time of periastron passage $T_{\rm peri}$, and $e$.

We have visually inspected the $O-C$ diagrams constructed as described previously,
in search of $O-C$ shapes allowed by Eq.~\ref{lteffect} (e.g., \citealt{ir59}).
In some cases, particularly when the period values found by \citet{so11} have turned out to be 
inadequate, we have constructed the $O-C$ diagrams with several different pulsation periods. 

Figure~\ref{method} shows one of the best candidates for a binary system containing an RRL star 
obtained in this way. In 
panel b of this figure, a preliminary orbital solution is also provided.
Figure~\ref{method}c demonstrates the accuracy of the fit;
the prewhitening for the orbital solution results in a LC which has an rms deviation of only 12~mmag.

A large fraction of stars show erratic
period changes, possibly connected to the Blazhko effect (\citealt{bl07}; see also \citealt{ju11}).
The Blazhko effect manifests itself as an amplitude and phase modulation of the LC of
RRL stars, and accordingly, variables affected by it have unreliable $O-C$ diagrams for our purposes.
For this reason, we have discarded binary candidates for which the LC shape and/or amplitude
showed correlated changes with the phase variations, which might be indicative of the Blazhko effect. 
However, the presence of the Blazhko effect is unclear in some of the variables, as blending with
another variable star and/or large photometric errors (in case of the fainter stars in the subsample),
together with the sparse coverage of the OGLE-III data, could mimic and/or mask the change
of the LC shape/amplitude of Blazhko variables.

\begin{figure*}
\includegraphics[angle=0,scale=.72]{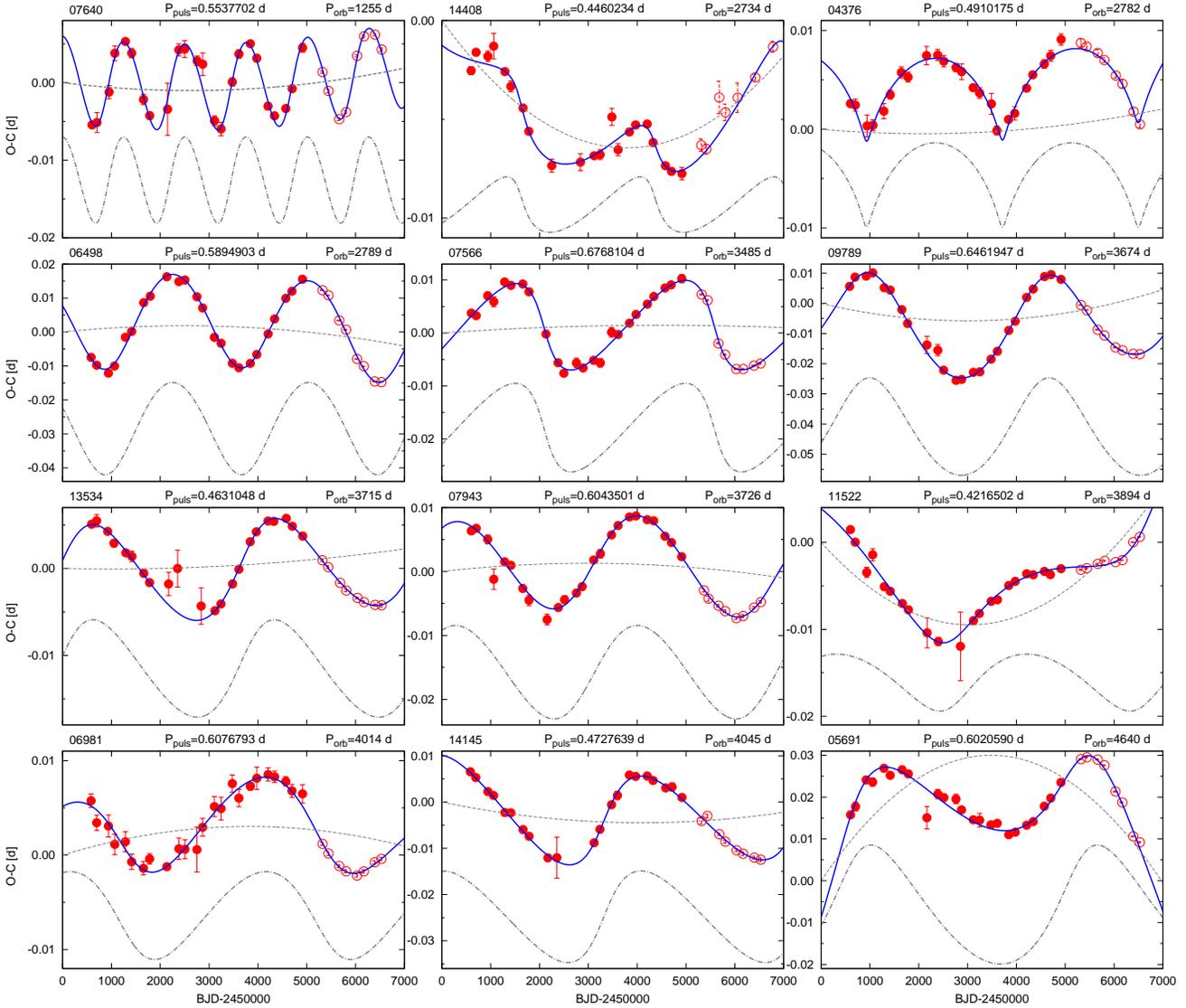}
\caption{$O-C$ diagrams of RRL binary candidates. The $O-C$ points ({\em filled circles}: OGLE-III data;
{\em empty circles}: OGLE-IV data) are fitted with the sum of a linear period change (Eq.~\ref{parabola},
{\em dashed line}) and the binary orbit (Eq.~\ref{lteffect}, {\em dot-dashed line}, shifted downwards for clarity).
For each star, the OGLE ID, pulsation period, and orbital period are given on top of the upper $x$-axis.
\label{results1}}
\end{figure*}

\subsection{Analysis of the candidates}

Following the above procedure, we have selected 29 potential binary candidates. 
In order to clarify the status of the stars in this sample, and also 
to refine the binary parameters of the candidates, we next combined the 
OGLE-III and IV datasets for these stars, thus increasing the baselines 
of their $O-C$ diagrams. The full analysis of the combined OGLE-III and IV 
datasets will be reported in a future paper.

\begin{table}
 \caption{OGLE IDs$^{a}$ of Uncertain RRL Binary Candidates}
 \label{nofit}
 \begin{tabular}{ccccc}
  \hline
  \multicolumn{2}{c}{Likely non-binaries} &&\multicolumn{2}{c}{Likely binaries with poor fits}\\
  \hline
  Symptom & ID && Symptom & ID\\
  \cline{1-2} \cline {4-5}\\
  Blazhko effect & 05075 && $P_{\rm orb}>6000$ d & 04522\\
                 & 07773 &&                      & 05135\\
                 & 10519 &&                      & 05152\\
                 & 12027 &&                      & 10891\\
                 & 13260 &&                      & 11683\\
                 & 13698 &&                      & 12611\\
                 &       &&                      & 12698\\
  \cline{1-2} \cline {4-5}\\
  Period changes & 05778 && Very small amplitude & 14852\\
                 & 06876 &&                      & \\
                 & 12195 &&                      & \\
  \hline
 \end{tabular}

 \medskip
 $^{a}$OGLE-BLG-RRLYR
\end{table}

\begin{table*}
 \caption{Fitted and derived parameters of the RRL binary candidates$^{a}$}
 \label{table_parameters}
 \begin{tabular}{cllcrccrcc}
 \hline
   ID & \multicolumn{1}{c}{$P_{\rm orb}$}&\multicolumn{1}{c}{$T_{\rm peri}$}& $e$  & \multicolumn{1}{c}{$\omega$} &
    $a \sin i$    &$\sigma$& \multicolumn{1}{c}{$\beta$}        & $K$ &$f(m)$\\
      & \multicolumn{1}{c}{(d)}  & \multicolumn{1}{c}{(d)} &  & \multicolumn{1}{c}{(deg)} & (AU) & (AU) & (d Myr$^{-1}$) &(km s$^{-1}$) & ($M_\odot$)\\
      \hline
07640 & $1255\pm  4$ & $9598\pm 54$ & $0.16\pm0.04$ & $ -30\pm15$ & $0.97\pm0.02$ & $0.16$ & $ 0.05\pm0.01$ & 8.5 & 0.0774\\
14408 & $2734\pm 38$ & $8871\pm 78$ & $0.60\pm0.12$ & $ 162\pm 8$ & $0.30\pm0.03$ & $0.11$ & $ 0.16\pm0.01$ & 1.5 & 0.0005\\
04376 & $2782\pm 19$ & $8147\pm 53$ & $0.79\pm0.11$ & $ -92\pm 6$ & $0.74\pm0.06$ & $0.11$ & $ 0.04\pm0.01$ & 4.7 & 0.0070\\
06498 & $2789\pm 18$ & $8137\pm134$ & $0.12\pm0.04$ & $ -82\pm16$ & $2.35\pm0.05$ & $0.16$ & $-0.11\pm0.03$ & 9.2 & 0.2228\\
07566 & $3485\pm 15$ & $8612\pm 33$ & $0.54\pm0.03$ & $-178\pm 3$ & $1.72\pm0.04$ & $0.14$ & $-0.03\pm0.02$ & 6.4 & 0.0556\\
09789 & $3674\pm 30$ & $7221\pm 81$ & $0.18\pm0.04$ & $  78\pm10$ & $2.80\pm0.06$ & $0.23$ & $ 0.24\pm0.03$ & 8.4 & 0.2162\\
13534 & $3715\pm 27$ & $6327\pm 73$ & $0.26\pm0.03$ & $  20\pm 5$ & $1.00\pm0.02$ & $0.22$ & $ 0.02\pm0.01$ & 3.0 & 0.0096\\
07943 & $3726\pm 40$ & $9273\pm150$ & $0.14\pm0.03$ & $ -15\pm13$ & $1.28\pm0.03$ & $0.20$ & $-0.05\pm0.02$ & 3.8 & 0.0200\\
11522 & $3894\pm153$ & $8751\pm411$ & $0.30\pm0.14$ & $ -61\pm24$ & $0.57\pm0.08$ & $0.12$ & $ 0.37\pm0.04$ & 1.7 & 0.0017\\
06981 & $4014\pm 85$ & $7349\pm206$ & $0.25\pm0.06$ & $-150\pm11$ & $0.82\pm0.04$ & $0.14$ & $-0.08\pm0.03$ & 2.3 & 0.0046\\
14145 & $4045\pm100$ & $9255\pm 81$ & $0.41\pm0.04$ & $  -3\pm 7$ & $1.88\pm0.08$ & $0.11$ & $ 0.10\pm0.02$ & 5.5 & 0.0541\\
05691 & $4640\pm119$ & $6010\pm314$ & $0.35\pm0.06$ & $  46\pm12$ & $2.54\pm0.17$ & $0.32$ & $-0.90\pm0.14$ & 6.3 & 0.1011\\
\hline
\multicolumn{10}{p{.9\textwidth}}{
$^{a}$ The columns, in order, correspond to the following quantities:
(1) OGLE ID, in the usual form OGLE-BLG-RRLYR plus the catalogue entry number;
(2) orbital period;
(3) time of periastron passage, in units of ${\rm BJD}-2440000$;
(4) eccentricity;
(5) argument of the periastron;
(6) projected semi-major axis;
(7) standard deviation of the fit $\sqrt{\mathrm{SSR}/[M-N]}$, where $\mathrm{SSR}$ is the sum of squared residuals, $N$
is the number of data points and $M$ is the number of free parameters;
(8) rate of change of the pulsation period;
(9) semi-amplitude of the radial velocity $K=2 \pi a \sin i /P_{\rm orb}\sqrt{1 - e^2}$;
(10) mass function $f(m)=a^3 \sin^3 i /P_{\rm orb}^2$, which is connected to the stellar masses through $f(m)=m_s^3 \sin^3 i /(m_{\rm RR} + m_s)^2$,
where $m_{\rm RR}$ is the mass of the RRL, and $m_s$ is the mass of the secondary.}
\end{tabular}
\end{table*}

We have combined the OGLE-III and IV datasets for each of our 29 candidates by correcting for the
difference between their average magnitudes, as given by \citet{so09} and \citet{so14}. 
We then repeated the previous analysis on the data thus combined. We have inspected
the new $O-C$ diagrams and the combined LCs, and have found strong evidence that 6 of the candidates
do in fact show the Blazhko effect, which was not evident from the relatively sparse OGLE-III data alone.
The $O-C$ variations of two additional candidates appear to be caused completely by their linear period changes,
while for a third star, it is probably caused by irregular period changes. These discarded candidates are listed
in the second column of Table~\ref{nofit}, leaving us with a refined sample of 20 binary candidates. 

Eq.~\ref{lteffect} 
does not take possible changes in the pulsation periods of the RRL stars into account. Such 
changes can significantly alter the $O-C$ diagram, even on relatively short ($\sim10$~yr) timescales. If the
period changes linearly with time, the $O-C$ diagram has the following shape: 

\begin{equation}
(O - C) (t)=c_0 + c_1 t + c_2 t^2, \label{parabola}
\end{equation}

\noindent where $c_2 \equiv \beta$ is the linear period-change rate. The increased time-base of the 
combined datasets allows us to take this effect fully into account. 
We thus fit the $O-C$ diagrams of the 20
remaining candidates with the sum of Eqs.~\ref{lteffect} and \ref{parabola}, and find that
reliable parameters can be derived for 12 of them. The other 8 variables either have orbital periods that are presumably
comparable or longer than the baseline of the available observations (therefore the parameters of the fit
are degenerate), or have very small $O-C$ amplitudes. These stars, which may still be 
bona-fide binaries and thus merit continued monitoring, are also listed in Table~\ref{nofit}. 

In order to derive the best possible parameters for this 12-star sample, 
we iterate the $O-C$ solution once, by means of an improved Fourier fit to a ($O-C$)-subtracted 
LC. We also $3\sigma$-clip the LCs, discarding 
the worst-quality (typically $\sim 1$ per cent) observations for each star. The
final $O-C$ points are determined by fitting these templates to the LC segments. 
The final diagrams, together with their fits, are shown in Figure~\ref{results1}.
Table~\ref{table_parameters} gives the fitted binary parameters for each star, 
as well as other relevant parameters of the fit.

\section{Discussion}

We have completed the first systematic search for binaries in a subsample of OGLE
Galactic bulge RRL stars utilizing the light-travel time effect.
Twenty probable binaries have been found analyzing the $O-C$ diagrams and LCs of
1952 OGLE-III bulge RRab variables, which represents about 1 per cent of this particular subsample
(fundamental-mode pulsation, $>$10~yr observational baseline). 
This allows us to very roughly constrain the RRL binary fraction, as follows. 

Approximately 50 per cent of RRab stars show the Blazhko effect \citep{ju09,ko10}.
Due to their LC phase and shape changes, determining their binarity through the $O-C$ method is
impossible. Therefore, all of the Blazhko binaries have necessarily been missed, and so the total 
binary fraction must be closer to 2 per cent than to 1 per cent. We assume that the fraction of stars whose 
binarity was missed due to erratic pulsation period changes is small in comparison. 
Binaries with very low inclinations and very long periods are also missed.
As we have no information about the fraction of long-period
binaries in the sample, and as our method is more sensitive to stars with high 
inclination (because the $O-C$ amplitude is proportional to $a \sin i$), we conservatively assume 
that at least half of the binaries are still missed due to these two selection effects. 
Based on these arguments, presumably at least 4~per cent of RRL variables reside in binary
systems in the sample. The recovery rate using the $O-C$ method is thus $\sim 25$~per cent, or perhaps lower. 

Very few RR~Lyrae stars have been reported to reside in binary systems yet (see Sect.~\ref{intro}).
The current sample allows us to assess the
observational requirements to discover additional such systems. The period distribution of the binary
candidates is highly skewed: we find no binaries with orbital periods shorter than $\sim3.5$~yr, and
there is a clustering of stars with orbital periods around $\sim10 - 11$~yr. Due to the long timescale, currently
OGLE is the only survey capable of discovering these variables, as high-precision ($0.01-0.02$~mag), 
long-baseline data are indispensable.
Indeed, the increased baseline achieved by incorporating the OGLE-IV data  
greatly improved the quality of the fits, as we were able to disentangle the effects of the
change in the RRL pulsation periods form the effect of binarity on the $O-C$ diagrams, thus 
leading to more robust binary parameters.
Note that we have succeeded in recovering RRLs with
Blazhko periods as short as $\sim0.5$~yr, which is close to the limit of period detection 
set by the data sampling~-- thus indicating that binary RRLs 
with similarly short periods would have been found through our analysis.  
The apparent lack of RRL variables in short-period binary
systems puts a tight constraint on the inclination (due to the wide orbits) of eclipsing
RRL binary systems, making our method much superior for the detection of
RRL binaries.

Their faintness (15-17~mag in $I$) makes the follow-up of our binary candidates
challenging. Still, two of the candidates (06498 and 09879) have \textit{higher} minimum masses
(calculated from the parameters in Table~\ref{table_parameters} for reasonable mass assumptions for
the RRL) than the RRL itself. The secondaries in some of these systems may be
stars which have evolved off from the main sequence, therefore bright enough to contribute significantly
to the light of the systems. If such a favorable case could be found, the mass ratio between the
two stars could be measured through the ratio of their orbital radial velocities.

At the distance of the bulge, the binary candidate with the longest major axis (09789, $\sim5.6$~AU)
would have a projected major axis of about 0.7~mas. Gaia astrometry will unfortunately be hampered by the high crowding and the faintness of the candidates. However, \citet{sa14} have demonstrated that $120 \, \mu$as astrometry with FORS2@VLT is attainable
on multi-year timescales, for targets with similar magnitudes as here.
A long-term astrometric follow-up program might thus be feasible, in order to determine the inclination
of these systems.  

In this study of a subsample of OGLE LCs we have demonstrated that RRL stars
can be detected in long-period binary systems, provided that high-quality, extended photometric datasets 
are used. We plan to extend our analysis to the whole OGLE bulge RRL dataset in the near future. 
Through long-term, dedicated monitoring using these techniques, and adequate follow-up, 
we will finally have a chance of determining the masses for a significant sample 
of RRL stars, and thereby directly constrain the theories of RRL star pulsation and evolution.

\section*{Acknowledgments}

G.H. acknowledges discussions with A. Tokovinin and K. He{\l}miniak. 
We also thank the referee for her/his helpful comments. 
Support for this project is provided by the Ministry for the Economy,
Development, and Tourism's Programa Iniciativa Cient\'{i}fica Milenio through grant
IC\,210009, awarded to the Millennium Institute of Astrophysics; by Proyecto Basal
PFB-06/2007; by Fondecyt grant \#1141141; and by CONICYT Anillo grant ACT\,1101.

\label{lastpage}


\begin{thebibliography}{99}

\bibitem[\protect\citeauthoryear{Benedict et al.}{2011}]{fbea11} Benedict G. F. et al., 2011, AJ, 142, 187

\bibitem[\protect\citeauthoryear{Bla\v{z}ko}{1907}]{bl07} Bla\v{z}ko S., 1907, AN, 175, 325

\bibitem[\protect\citeauthoryear{Bono et al.}{1996}]{gbea96} Bono G., Caputo F., Castellani V., Marconi M., 1996, ApJ, 471, L33 

\bibitem[\protect\citeauthoryear{Cacciari}{2013}]{cc13} Cacciari C., 2013, in R. de Grijs, ed., IAU Symp. 289, Advancing the Physics of Cosmic Distances, p. 101 

\bibitem[\protect\citeauthoryear{Catelan}{1992}]{mc92} Catelan M., 1992, A\&A, 261, 457 

\bibitem[\protect\citeauthoryear{Catelan}{2009}]{mc09} Catelan M., 2009, Ap\&SS, 320, 261

\bibitem[\protect\citeauthoryear{Catelan \& Cort{\'e}s}{2008}]{cc08} Catelan M., Cort{\'e}s C., 2008, ApJ, 676, L135

\bibitem[\protect\citeauthoryear{Catelan \& de Freitas Pacheco}{1993}]{cdf93} Catelan M., de Freitas Pacheco J. A., 1993, AJ, 106, 1858

\bibitem[\protect\citeauthoryear{Dambis et al.}{2013}]{da13} Dambis A. K., Berdnikov L. N., Kniazev A. Y., Kravtsov V. V., Rastoguev A. S., Sefako R., Vozyakova O. V., 2013, MNRAS, 435, 3206

\bibitem[\protect\citeauthoryear{D\'ek\'any et al.}{2008}]{de08} D\'ek\'any I. et al., 2008, MNRAS, 386, 521

\bibitem[\protect\citeauthoryear{D\'ek\'any et al.}{2013}]{de13} D\'ek\'any I., Minniti D., Catelan M., Zoccali M., Saito R. K., Hempel M.,
Gonzalez O. A., 2013, ApJ, 776, L19

\bibitem[\protect\citeauthoryear{Drake et al.}{2013a}]{ajdea13a} Drake A. J. et al., 2013a, ApJ, 763, 32

\bibitem[\protect\citeauthoryear{Drake et al.}{2013b}]{ajdea13b} Drake A. J. et al., 2013b, ApJ, 765, 154

\bibitem[\protect\citeauthoryear{Eastman et al.}{2010}]{esg10} Eastman J., Siverd R., Gaudi B. S., 2010, PASP, 122, 935

\bibitem[\protect\citeauthoryear{Feast et al.}{2008}]{mfea08} Feast M. W., Laney C. D., Kinman T. D., van Leeuwen F., Whitelock P. A., 2008, MNRAS, 386, 2115

\bibitem[\protect\citeauthoryear{Gieren et al.}{2014}]{wgea14} Gieren W. et al., 2014, ApJ, 786, 80 

\bibitem[\protect\citeauthoryear{Guggenberger \& Steixner}{2014}]{gs14} Guggenberger E., Steixner J., 2014, arXiv:1411.1555v1

\bibitem[\protect\citeauthoryear{Hertzsprung}{1919}]{he19} Hertzsprung E., 1919, AN, 210, 17

\bibitem[\protect\citeauthoryear{Irwin}{1952}]{ir52} Irwin J. B., 1952, ApJ, 116, 211

\bibitem[\protect\citeauthoryear{Irwin}{1959}]{ir59} Irwin J. B., 1959, AJ, 64, 149

\bibitem[\protect\citeauthoryear{Jurcsik et al.}{2009}]{ju09} Jurcsik J. et al., 2009, MNRAS, 400, 1006

\bibitem[\protect\citeauthoryear{Jurcsik et al.}{2011}]{ju11} Jurcsik J., Szeidl B., Clement C., Hurta Zs., Lovas M., 2011, MNRAS, 411, 1763

\bibitem[\protect\citeauthoryear{Kennedy et al.}{2014}]{ke14} Kennedy C. R. et al., 2014, ApJ, 787, 6

\bibitem[\protect\citeauthoryear{Kolenberg et al.}{2010}]{ko10} Kolenberg K. et al. 2010, ApJ, 713, L198

\bibitem[\protect\citeauthoryear{Lee}{1992}]{ywl92} Lee Y.-W., 1992, AJ, 104, 1780

\bibitem[\protect\citeauthoryear{Li \& Qian}{2014}]{lq14} Li L.-J., Qian S.-B., 2014, MNRAS, 444, 600

\bibitem[\protect\citeauthoryear{Petersen}{1973}]{jop73} Petersen J. O., 1973, A\&A, 27, 89 

\bibitem[\protect\citeauthoryear{Pietrzy{\'n}ski et al.}{2010}]{gpea10} Pietrzy{\'n}ski G. et al., 2010, Nature, 468, 542 

\bibitem[\protect\citeauthoryear{Pietrzy{\'n}ski et al.}{2012}]{pi12} Pietrzy\'nski G. et al., 2012, Nature, 484, 75

\bibitem[\protect\citeauthoryear{Popielski et al.}{2000}]{bpea00} Popielski B. L., Dziembowski W. A., Cassisi S., 2000, AcA, 50, 491

\bibitem[\protect\citeauthoryear{Pr\v{s}a et al.}{2008}]{pr08} Pr\v{s}a A., Guinan E. F., Devinney E. J., Engle S. G., 2008, A\&A, 489, 1209

\bibitem[\protect\citeauthoryear{Saha \& White}{1990}]{sw90} Saha A., White R.~E., 1990, PASP, 102, 148; erratum: PASP, 102, 495 

\bibitem[\protect\citeauthoryear{Sahlmann et al.}{2014}]{sa14} Sahlmann J., Lazorenko P. F., S\'egransan D., Mart\'in E. L., Mayor M., Queloz D., Udry, S., 2014, A\&A, 565, 20

\bibitem[\protect\citeauthoryear{Sandage}{2006}]{as06} Sandage A., 2006, AJ, 131, 1750 

\bibitem[\protect\citeauthoryear{Sesar et al.}{2013}]{se13} Sesar B. et al., 2013, AJ, 146, 21

\bibitem[\protect\citeauthoryear{Smolec et al.}{2013}]{sm13} Smolec R. et al., 2013, MNRAS, 428, 3034

\bibitem[\protect\citeauthoryear{S{\'o}dor}{2012}]{so12} S\'odor {\'A}., 2012, Konkoly Observatory Occasional Technical Notes, 15, 1

\bibitem[\protect\citeauthoryear{Soszy\'nski et al.}{2003}]{so03} Soszy\'nski I. et al., 2003, AcA, 53, 93

\bibitem[\protect\citeauthoryear{Soszy\'nski et al.}{2009}]{so09} Soszy\'nski I. et al., 2009, AcA, 59, 1

\bibitem[\protect\citeauthoryear{Soszy\'nski et al.}{2011}]{so11} Soszy\'nski I. et al., 2011, AcA, 61, 1

\bibitem[\protect\citeauthoryear{Soszy\'nski et al.}{2014}]{so14} Soszy\'nski I. et al., 2014, AcA, 64, 177

\bibitem[\protect\citeauthoryear{Stancliffe et al.}{2013}]{st13} Stancliffe R. J., Kennedy, Lau H. H. B., Beers T. C., 2013, MNRAS, 435, 698

\bibitem[\protect\citeauthoryear{Sterken}{2005}]{st05} Sterken C., 2005, in Sterken C., ed., ASP Conf. Ser. Vol. 335,
p. 3

\bibitem[\protect\citeauthoryear{Szabados}{2003}]{sza03} Szabados L., 2003, IBVS, 5394, 1

\bibitem[\protect\citeauthoryear{Torrealba et al.}{2015}]{gtea14} Torrealba G. et al., 2015, MNRAS, 446, 2251

\bibitem[\protect\citeauthoryear{Wade et al.}{1999}]{wa99} Wade R. A., Donley J., Fried R., White R. E., Saha A., 1999, AJ, 118, 2442


\end{thebibliography}
\end{document}